\documentstyle{article}
\def\1ad{\mbox{\normalsize $^1$}}
\def\2ad{\mbox{\normalsize $^2$}}
\def\3ad{\mbox{\normalsize $^3$}}
\def\4ad{\mbox{\normalsize $^4$}}
\def\5ad{\mbox{\normalsize $^5$}}
\def\6ad{\mbox{\normalsize $^6$}}
\def\7ad{\mbox{\normalsize $^7$}}
\def\8ad{\mbox{\normalsize $^8$}}
\def\makefront{\vspace*{1cm}\begin{center}
\def\newtitleline{\\ \vskip 5pt}
{\Large\bf\titleline}\\
\vskip 1truecm
{\large\bf\authors}\\
\vskip 5truemm
\addresses
\end{center}
\vskip 1truecm
{\bf Abstract:}
\abstracttext
\vskip 1truecm}
\newcommand{\ft}[2]{{\textstyle\frac{#1}{#2}}}

\def\e{{\rm e}}
\def\d{{\rm d}}
\begin {document}
\def\titleline{
Examples of stationary BPS solutions \newtitleline
in $N=2$ super\-gravity theories with $R^2$-interactions
}
\def\authors{
G.L. Cardoso\1ad, B. de Wit\2ad, 
J. K\"appeli\2ad, T. Mohaupt\3ad
}
\def\addresses{
  \1ad Humboldt Universit\"at zu Berlin, Berlin, Germany \\ \2ad Institute for
  Theoretical Physics and Spinoza Institute,\\ Utrecht University, Utrecht,
  The Netherlands \\ \3ad Physics Department, Stanford University, Stanford,
  USA
}
\def\abstracttext{
  We discuss explicit examples of BPS solutions in four-dimensional $N=2$
  supergravity with $R^2$-interactions. We demonstrate how 
  to construct solutions by iteration. Generically, the presence
  of higher-curvature interactions leads to non-static spacetime line
  elements. We comment on the existence of horizons for 
  multi-centered solutions.
}
\begin{flushright}
HU-EP-00/56\\[-0.25mm]
ITP-UU-00/46\\[-0.25mm]
SPIN-2000/35\\[-0.25mm]
SU-ITP 00/34\\[-0.25mm]
{\tt hep-th/0012232}\\[5mm]
December 2000
\end{flushright}
\large
\makefront
\section{Introduction}
This note is a follow-up of a recent study \cite{sept} of stationary BPS
solutions in four-di\-men\-sional $N=2$ supergravity theories with
higher-curvature interactions and with an arbitrary number of abelian
vector multiplets and neutral hypermultiplets. Working within the
framework of the superconformal multiplet calculus an analysis of
solutions with $N=2$ and 
residual $N=1$ supersymmetry was presented. It was established that
there exist only two classes of fully supersymmetric vacua, namely
Bertotti-Robinson and Minkowski spacetime, and that the corresponding
solutions are fully specified by the charges carried by the field
configurations. As a consequence of the uniqueness of the horizon
geometry the so-called fixed-point behavior 
\cite{ferkalstr} is thus established in the presence of $R^2$-terms.
Furthermore, concise equations that govern stationary BPS solutions were
derived. In the absence of $R^2$-interactions our conclusions agree with
previous results (here we refer in particular to the recent work
\cite{denef1}). In 
this note we demonstrate that the construction of BPS solutions in
$N=2$ supergravity with higher-curvature interactions is in general
very complicated and show, by studying a few explicit examples, how to
proceed iteratively. We address the fact that  multi-centered BPS 
solutions are non-static due to 
the presence of $R^2$-interactions, and present arguments why these
configurations may still have regular horizons at the charge centers.
\section{Residual supersymmetry}
For what follows we adopt the definitions and conventions used in \cite{sept}.
The line element of a stationary solution is parametrized by
\begin{equation}
  \label{eq:lineelementdef}
  \d s^2 = - \e^{2g} \left(\d t + \sigma_m \d x^m \right)^2 + \e^{-2g}\, g_{mn}\d x^m
  \, \d x^n\,.  
\end{equation}
Imposing residual $N=1$ supersymmetry subject to the embedding condition
\begin{equation}
  \label{eq:embedd}
  h \, \epsilon_i = \varepsilon_{ij} \, \gamma_0 \, \epsilon^j \,,
\end{equation}
where $h$ is a phase, leads to stringent restrictions. For instance, $g_{mn}$
needs to be a flat three-dimensional metric and the hypermultiplet scalars
must be constant. To transparently present the restrictions of residual
supersymmetry on the vector multiplet sector it is advantageous to rescale the
vector multiplet scalars and to express the Lagrangian in terms of the scale
and U(1) invariant variables $Y^I$ ($I=0,\ldots, n $), which are shown to be
expressed in terms of harmonic functions $H^I(\vec x)$ and $H_I(\vec x)$ according to
\begin{equation} \label{eq:gen-stab-eqs}
    \pmatrix{Y^I-{\bar Y}^I\cr\noalign{\vskip2mm} F_I(Y,\Upsilon)-{\bar
    F}_I(\bar Y,\bar \Upsilon) } = i
    \pmatrix{H^I\cr\noalign{\vskip2mm}H_I}\,. 
\end{equation}
These $2(n+1)$ equations for the real and imaginary parts of the scalars $Y^I$
constitute the so-called generalized stabilization equations. In the absence
of $R^2$-terms these equations were first conjectured in \cite{BLS}.  Here
$F_I(Y,\Upsilon)$ is the partial derivative with respect to $Y^I$ of a
homogeneous holomorphic function $F(Y, \Upsilon)$. This function characterizes
the couplings of the vector multiplets. The field $\Upsilon$ appearing in
$F(Y, \Upsilon)$ as an extra holomorphic parameter is the (rescaled) scalar of
a chiral background superfield which, upon identification with the square of
the Weyl superfield, parametrizes the $R^2$-interactions. The harmonic
functions are related to the electric and magnetic charges carried by the
field configurations. In the case of multiple centers they are superpositions
of the harmonic functions associated with the electric ($q_{AI}$) and magnetic
($p^I_{\,A}$) charges carried by the centers located at $\vec{x}_A$,
\begin{equation}
  H^I(\vec x) = h^I + \sum_A \frac{p_A^I}{|\vec{x} - \vec{x}_A|} \;,\qquad
H_I(\vec x) = h_I + \sum_A \frac{q_{AI}}{|\vec{x} - \vec{x}_A|} \;, 
\end{equation}
where the $(h^I,h_J)$ are constants.  It is remarkable that apart from
(\ref{eq:gen-stab-eqs}) the only other equations that must be solved to fully
specify the stationary BPS solutions, are the ones for the spacetime line
element,
\begin{eqnarray}\label{eq:line}
  i\,  \Big[{\bar Y}^I F_I (Y, \Upsilon)
- {\bar F}_I ({\bar Y}, {\bar \Upsilon}) {Y}^I\Big] + \ft12\, \chi\, \e^{-2g} & =
&  128 i \,{\rm e}^{g} \,\nabla^p
\Big[\left(\nabla_p {\rm e}^{-g}\right)\,(F_\Upsilon-\bar F_\Upsilon)\Big] \nonumber
\\[1mm] &&
- 32 i \,{\rm e}^{4g}
\,(R(\sigma)_p)^2 (F_\Upsilon-\bar F_\Upsilon)
\nonumber \\[1mm] && -64 \,{\rm e}^{2g} R(\sigma)_p\,\nabla^p  (F_\Upsilon+\bar F_\Upsilon ) \,,
\nonumber \\[2mm]
H^I \stackrel{\leftrightarrow}\nabla_p H_I
 +\ft12 \, \chi  \, R(\sigma)_p &=& - 128i \,  \nabla^q \Big[
\nabla_{[p}  \Big( {\rm e}^{2g} \, R(\sigma)_{q]}\,(F_\Upsilon  - {\bar
F}_\Upsilon)   \Big) \Big]
 \nonumber \\[1mm] &&    - 128 \,  \nabla^q \Big[ 2 \nabla_{[p} g \,\nabla_{q]} (F_\Upsilon +
{\bar F}_\Upsilon)\Big]
\,, \label{ak2}
\end{eqnarray}
where $\Upsilon$ is constrained to take the form 
\begin{equation}\label{eq:ups}
\Upsilon= - 64 \, \Big(\nabla_p g - \ft12 i {\rm e}^{2g}\,
R(\sigma)_p\Big)^2\,. 
\end{equation}
Here $p$, $q$ denote three-dimensional tangent space indices and
$R(\sigma)_p$ is the dual of the field strength of $\sigma$,
$R(\sigma)^p=\varepsilon^{pqs} \nabla_q \sigma_s$. Furthermore,
$F_\Upsilon = \partial_{\, \Upsilon} F(Y,\Upsilon)$ and $\chi$ denotes
a negative constant which fixes the scale. 
All other fields, such as the field strengths of the vector multiplets, are
given by explicit formulae in terms of $g$, $R(\sigma)_p$, and the moduli
fields $Y^I$. Solutions to the above equations automatically satisfy all
equations of motion.
\section{Examples of BPS solutions}
The equations (\ref{eq:gen-stab-eqs}), (\ref{eq:line}), and
(\ref{eq:ups}) that determine stationary BPS solutions can be solved
explicitly only in very few cases. For instance, it is often difficult
to obtain the solution to the generalized stabilization equations
(\ref{eq:gen-stab-eqs}) in closed form, even in the absence of
$R^2$-terms.  Furthermore, equations (\ref{eq:line}) and
(\ref{eq:ups}) are coupled and can usually be solved only 
iteratively. For concreteness, let us consider a first simple example, 
\begin{equation}
  \label{eq:quadr1}
  F(Y,\Upsilon) = -\ft{1}{2}\,i\,  Y^I\, \eta_{IJ} \, Y^J + c\, \Upsilon\,, 
\end{equation}
where $\eta$ is a real symmetric matrix and $c$ a complex number. In
this case it is simple to solve the equations
(\ref{eq:gen-stab-eqs}),
\begin{equation}
  \label{eq:stabsol1}
  Y^I = \ft12 \left( i H^I - \eta^{IJ} H_J \right)\,,\quad 
  F_I = \ft12 \left( i H_I + \eta_{IJ} H^J  \right)\,,
\end{equation}
where $\eta_{IJ} \eta^{JK} = \delta_{I}^{\ K}$.  The dependence on the
$R^2$-background will enter only when solving for the line element, as
we will discuss shortly. This is rather the exception than the
rule. Consider, for instance, the coupling function which arises in
Calabi-Yau threefold compactifications in the large-volume limit,
\begin{equation}\label{eq:cyprep}
  F(Y,\Upsilon) =\frac{D_{ABC}\, Y^A \, Y^B \, Y^C}{Y^0} + d_A\,
\frac{ Y^{A} }{Y^0}\, \Upsilon  
\,,
\end{equation}
with $ A,B,C = 1,2, \ldots, n $. We construct solutions
to this model satisfying $H^0=0$ so that  $Y^0$ is real.
Introducing the matrix $ D_{AB} = D_{ABC} H^C$ and assuming its
invertibility $D_{AB} D^{BC} = \delta_{A}^{\ C}$, the stabilization
equations can be solved to all orders in $\Upsilon$,
\begin{eqnarray}\label{eq:cysol}
   Y^A & = &  \ft{1}{6} Y^0 D^{AB} \left( H_B+ i d_B
 \frac{\Upsilon-\bar\Upsilon}{Y^{0}}\right) + \ft{1}{2}\, i\,
 H^A\,,\nonumber \\ 
(Y^0)^2 &  = &\frac{D_{ABC} H^A H^B H^C - \frac{1}{3} (\Upsilon-\bar
\Upsilon)^2
  d_A D^{AB} d_B - 2 (\Upsilon+ \bar\Upsilon) d_A H^A }{4 \left( H_0 +
\frac{1}{12}H_A D^{AB} H_B \right)}\;.
\end{eqnarray}
For more complicated $F(Y,\Upsilon)$, solving (\ref{eq:gen-stab-eqs}) 
can become very involved. It may be possible to cast $F(Y,\Upsilon)$
into a power series expansion (possibly after an electric/magnetic
duality transformation) in which case the generalized stabilization
equations (\ref{eq:gen-stab-eqs}) can be solved iteratively in powers
of $\Upsilon$, 
\begin{eqnarray}
  \label{eq:expansions}
   Y^I(H,\Upsilon,\bar \Upsilon) & = & \sum_{n,m} Y^I_{(n,m)}(H) \, 
  \Upsilon^n \bar\Upsilon ^m\,.
\end{eqnarray}
Clearly, $F_I(Y,\Upsilon)$ and $F_\Upsilon(Y,\Upsilon)$ will have
corresponding expansions in $\Upsilon$ and $\bar \Upsilon$ once the
solutions $Y^I(H, \Upsilon,\bar\Upsilon)$ are inserted, and one could
in principle, by treating $\Upsilon$ as a formal expansion parameter,
proceed to solve (\ref{eq:line}) and (\ref{eq:ups}). Since such a
procedure is not feasible in practice, the question arises whether it
makes sense to solve the equations (\ref{eq:line}) and (\ref{eq:ups})
iteratively and to truncate at some suitable order. To address this
question we recall that the function $F(Y,\Upsilon)$ is homogeneous of
degree two, 
\begin{equation}
F(\lambda Y, \lambda^2\, \Upsilon) = \lambda^2\, F(Y,\Upsilon)\,.
\end{equation}
It follows that 
\begin{equation}
   Y^I F_I(Y,\Upsilon) + 2 \Upsilon \,F_\Upsilon(Y,\Upsilon) = 2
   F(Y,\Upsilon)\,, 
\end{equation}
and in particular we have $F_I(\lambda Y, \lambda^2\, \Upsilon) =
\lambda\,F_I(Y,\Upsilon)$. This shows that the equations
(\ref{eq:gen-stab-eqs}) are invariant under this rescaling if we let
the harmonic functions $H^I$ and $H_I$ scale with weight
one. Therefore the coefficient functions $Y^I_{(n,m)}(H)$ in
the expansion (\ref{eq:expansions}) will scale with weight $1-2(n+m)$,
such that every power of $\Upsilon$ is accompanied by a net amount of
two inverse powers of harmonic functions $H^{-2}$. The expressions
(\ref{eq:cysol}) illustrate this feature.  In a similar way
homogeneity is reflected at the level of the Lagrangian. Therefore,
corrections due to $R^2$-interactions become subleading whenever
$|\Upsilon|\ll H^2$.  One can pinpoint such a hierarchy when one
encounters supersymmetry enhancement in the neighborhood of a charged
center.  There, $|\Upsilon|$ has a $1/r^2$-fall-off proportional to a
charge-independent constant, while the harmonic functions fall off as
$Q/r$, where $Q$ is the charge carried by the center. This is the
reason why the corrections to the entropy of BPS black holes
\cite{dez}, for instance, are subleading in the limit 
of large charges. In fact, owing to homogeneity, the entropy will have 
an expansion of the form ${\cal S} = \pi\, \sum_{n\geq 0} {\cal
S}^{(n)} Q^{2-2n}$,  where the coefficients ${\cal S}^{(n)}$ are
independent of the charges.  According to the arguments presented
above, homogeneity implies that the expansion of the line element
takes the schematic form  
\begin{equation}\label{eq:chargeex}
 \e^{-2g} \sim \sum_{n\geq 0} \alpha_{(n)} |\Upsilon|^n \,H^{2-2n} \,,
 \quad R(\sigma) \sim 
 \sum_{n\geq 0} \beta_{(n)} |\Upsilon|^n\, H^{2-2n}\,,
\end{equation}
where $\alpha_{(n)}$ and $\beta_{(n)}$ are independent of the harmonic
functions. This shows that a truncation at some finite order may be
sensible only in situations in which $\Upsilon$ falls off much
stronger than the harmonic functions $H^2$ when moving away from the
centers.  

Let us reconsider the holomorphic coupling function
(\ref{eq:quadr1}). In this simple example we can use $c$ as an
expansion parameter, since by homogeneity every power of $c$ will
always be accompanied by two inverse powers of harmonic
functions. After solving the generalized stabilization equations
(\ref{eq:stabsol1}) the remaining equations (\ref{eq:line}) reduce to 
\begin{eqnarray}\label{eq:theeq}
   \left(H^I \eta_{IJ} H^J + H_I \eta^{IJ} H_J\right)  
+ \chi\, \e^{-2g}  &=&  256\,  i \, (c-\bar c)\, \Big[ 
\e^g \nabla_p^2 \e^{-g}  -
(\ft12 {\rm e}^{2g} R(\sigma)_p)^2 \Big]
\,,
 \nonumber \\
 2 H^I \stackrel{\leftrightarrow}\nabla_p H_I +  \chi  \, R(\sigma)_p 
&=& 
   - 256\,  i \, (c-\bar c) \,  \nabla^q \Big[
\nabla_{[p}  \Big( {\rm e}^{2g} \, R(\sigma)_{q]}\,\Big) \Big]
\label{eq:eqns} \,.
\end{eqnarray}
The case where $c$ is real corresponds to adding a total derivative
term to the action. Above formulae show that the line element stays
unaltered in this case, while $\Upsilon$ is given by 
\begin{equation}
  \Upsilon  =  64 \left[\frac{ \left(H^I - i \eta^{IJ} H_J
\right)\nabla_p H_I - \left( H_I +i \eta_{IJ}H^J\right) \nabla_p H^I}{
H^I \eta_{IJ} H^J + H_I \eta^{IJ} H_J }\right]^2\,. 
\end{equation}
There are other less obvious situations where the dependency on $c$
drops out of the equations (\ref{eq:theeq}). This is the case, for
instance, when $R(\sigma)_p=0$ and all the harmonic functions are
proportional to $\e^{-g}$,  
  \begin{equation}
    H^I =  a^I\, \e^{-g} \,,\quad H_I = a_I\, \e^{-g}\,,
  \end{equation}
where $(a^I, a_I)$ are constants.
  
Generically, however, the right-hand sides of (\ref{eq:theeq}) do not
vanish and one has to rely on an iterative approach.  For mutually
local charges, $H^I \stackrel{\leftrightarrow}\nabla_p H_I = 0$,
static solutions with $R(\sigma)_p=0$ are possible. This is what we
want to investigate in the following. The remaining equation, 
\begin{equation}
  \label{eq:theremeq}
\e^{-2g} \Big( \chi -  256\,i (c-\bar c)\, 
\e^{3g} \nabla_p^2 \e^{-g}\Big)  = -\left(H^I \eta_{IJ} H^J + H_I
\eta^{IJ} H_J\right) \,, 
\end{equation}
is a non-linear differential equation for $\e^{-g}$. To zeroth order
in $c$ we find
\begin{equation}
  [\e^{-2g}]^{(0)} =  -\chi^{-1}\,\left(H^I \eta_{IJ} H^J + H_I \eta^{IJ}
   H_J\right)\,.
\end{equation}
Making the ansatz $\e^{-g} = \sum_{n\geq 0} [\e^{-g}]^{(n)}\big(256 i
(c-\bar c)/\chi\big)^n$ the
line element is determined iteratively by  
\begin{equation}\label{eq:etogit}
  [\e^{-g}]^{(n)} =\ft12 \, [\e^{2g}]^{(0)} \, \Big( \nabla_p^2\,
[\e^{-g}]^{(n-1)}  
- \sum_{i,j,k}{}^{'}  [\e^{-g}]^{(i)} [\e^{-g}]^{(j)}
[\e^{-g}]^{(k)}\Big)\,, 
\end{equation}
where the truncated sum $\sum{}^{\prime}_{i,j,k}$ runs over all $0\leq
i,j,k < n $ subject to $i+j+k = n $.  The presence of the overall
factor $ [\e^{2g}]^{(0)} \sim H^{-2}$ on the right-hand side of
(\ref{eq:etogit}) indeed induces the expansion indicated by
(\ref{eq:chargeex}). 

Let us return to the more complicated example (\ref{eq:cyprep}). In
this case we can use the $d_A$ as expansion parameters. The exact 
solution to the generalized stabilization equations for the case
$H^0=0$ are given in (\ref{eq:cysol}). We find by direct calculation
that   
\begin{equation}\label{eq:YFex}
i \Big[ {\bar Y}^I F_I - Y^I {\bar F}_I \Big]   = 
 - \frac{D_{ABC}
H^A H^B H^C}{Y^0}  + d_A H^A \frac{ \Upsilon + \bar \Upsilon}{Y^0}\,,
\end{equation}
and 
\begin{equation}\label{eq:Fupsex}
  F_{\Upsilon}- {\bar F}_{{\Upsilon}}  =  
\frac{i d_A H^A}{ Y^0} 
\,,\quad  F_{\Upsilon}+ {\bar F}_{{\Upsilon}} = \ft{1}{3}\, d_A D^{AB}
\left( H_B + i d_B \frac{\Upsilon-\bar\Upsilon}{Y^0}\right)\,. 
\end{equation}
To zeroth order in $d_A$ the solutions to (\ref{eq:line}) are given by  
\begin{equation}
  \label{eq:leading}
 \ft12\,\chi\, \e^{-2g}   =  \frac{D_{ABC}
H^A H^B H^C}{ [Y^0]^{(0)}} +{\cal O}(d_A)\nonumber \,, \quad  
\ft12\, \chi\,R(\sigma)_p   =   - H^I 
\stackrel{ \leftrightarrow} \nabla_p H_I +{\cal O}(d_A) \,,
\end{equation}
where $[Y^0]^{(0)}$ is the zeroth order expression for $Y^0$,
\begin{equation}
  ([Y^0]^{(0)})^2  = \frac{D_{ABC} H^A H^B H^C}{4 (H_0 + \ft{1}{12}
 H_A D^{AB} H_B )} \, .
\end{equation}
Keeping track of the terms coming from (\ref{eq:Fupsex}) the
expressions for the line element up to second order in $d_A$ are
readily obtained from (\ref{eq:line}),  
\begin{eqnarray}
\ft{1}{2} \, \chi\, {\rm e}^{-2g} &=& \frac{D_{ABC}
H^A H^B H^C}{ [Y^0]^{(0)}} - 128 \gamma^{-1} \nabla^p \left[
(\nabla_p \gamma)  \,  \frac{d_A H^A }{ [Y^0]^{(0)}} \right]
 + 32  \gamma^{-4} 
(\rho_p)^2 \frac{d_A H^A }{ [Y^0]^{(0)}}  \nonumber\\
& & - \ft{64}{3} \gamma^{-2}\, \rho_p \nabla^p
 \Big( d_A D^{AB} H_B \Big) + {\cal O}(d_A\, d_B)\,, \nonumber\\
\ft12\,\chi\,R(\sigma)_p &=& - H^I 
\stackrel{ \leftrightarrow} \nabla_p H_I+\frac{256}{3}
 \nabla^q \Big[ \gamma^{-1}
(\nabla_{[p} \gamma)  \nabla_{q]} (d_A D^{AB} H_B) \Big] \nonumber\\
&& + 128  \nabla^q\nabla_{[p} \left( \gamma^{-2} \rho_{q]}
\, \frac{d_A H^A }{ [Y^0]^{(0)}}\right) + {\cal O}(d_A\, d_B) \,,
\end{eqnarray}
where $\gamma^2$ and $\rho_p$ are abbreviations for the zeroth order
results as given in (\ref{eq:leading}) for $\e^{-2g}$ and
$R(\sigma)_p$, respectively. We see again that the first-order
approximation to the line element is cast into the 
form (\ref{eq:chargeex}), the first correction
from the $R^2$-terms being suppressed by one power of $H^{-2}$.

\section{Static versus stationary BPS solutions}
As pointed out in \cite{sept}, higher-curvature interactions induce
non-static pieces in the line element even for BPS configurations
with mutually local charges, $H^I \stackrel{ \leftrightarrow} \nabla_p
H_I = 0$. This effect can be observed in the previously discussed
example, but it does not arise when the solution has only one
center because the right-hand side of the second equation
(\ref{eq:line}) vanishes due to rotational symmetry. Therefore the
question arises whether $R^2$-interactions still allow for regular
multi-centered 
black hole solutions. Again, it is difficult to address this question in
full generality.  Due to (\ref{eq:chargeex}) we expect $\e^{-2g}$
generically to diverge as $ |\vec x - \vec x_A|^{-2}$ as one
approaches one of the charge centers $\vec{x}_A$. When the charges are
not mutually local one expects $R(\sigma)_p$  to behave as $|\vec x -
\vec x_A|^{-3}$. For mutually local charges, on the other hand, the
singularity of $R(\sigma)_p$ is, in fact, milder.  Inspection of the
expressions for the curvature components given in \cite{sept} shows
that every term involving $R(\sigma)_p$ is accompanied by a sufficient
amount of factors of $\e^{g}$ such that the effect of the non-static
pieces of the line element disappears near the centers. This 
indicates that multi-centered BPS solutions still possess an
${\rm AdS}_2 \times S^2$  geometry there. The following example
illustrates this. 

Let us consider a simple model describing pure supergravity with the
particular $R^2$-interactions given by
\begin{equation}
  F(Y,\Upsilon)  = - \ft{1}{2}\, i \,  (Y^0)^2 + b \,
  \frac{\Upsilon^2}{(Y^0)^2}\,\, .   
\end{equation}
We assume $b$ to be a real constant. By the same arguments as given in
the previous section, we use $b$ as an expansion parameter.  We solve  
the generalized stabilization equations for the purely electric
situation $H^0=0$, $H_0\equiv H$. To zeroth order in $b$ we find
\begin{equation}
  \e^{-2g} = - \chi^{-1}\,{H^2}+{\cal O}(b)\,, \quad R(\sigma)_p = {\cal
  O}(b)\,,\quad 
  \Upsilon = -64\, H^{-2}\,{(\nabla_p H)^2} + {\cal O}(b)\,.
\end{equation}
To calculate the next-order corrections to the line element one needs 
$F_\Upsilon+{\bar F}_{\Upsilon}$ to first order in $b$,
\begin{equation}\label{eq:Fups1}
  F_\Upsilon+\bar{F}_{\Upsilon}= -2^{10} \, b\, H^{-4}\, (\nabla_p H)^2
  + {\cal O}(b^2)\,. 
\end{equation}
We consider a harmonic function $H$ with multiple centers located at
$\vec{x}_A$. For simplicity let us assume that center $A=1$ is at 
$\vec{x}_1 =0$ and calculate $R(\sigma)_p$ around this
center. Therefore we expand  the harmonic function appearing in above
expression in powers of  $|\vec{x}|$,
\begin{eqnarray}\label{eq:exph}
H &=& h + \sum_A \frac{q_A}{|\vec{x} - \vec{x}_A|} =
\frac{q_1}{|\vec{x}|} + h  + \sum_{A \neq 1} \frac{q_A}{|\vec{x}_A|} +
{\cal O}(|\vec{x}|)  \,, \nonumber\\ 
\nabla_p H &=& - q_1 \frac{x_p}{|\vec{x}|^3} + \sum_{A \neq 1} q_A 
\frac{x_{Ap}}{|\vec{x}_A|^3} + {\cal O}(|\vec{x}|) \,, \nonumber\\
\nabla_p \nabla_q H &=& q_1 \left( 3 \frac{x_p\, x_q}{|\vec{x}|^5} 
- \frac{\delta_{pq}}{|\vec{x}|^3} \right) + \sum_{A \neq 1}q_A \left( 3
\frac{x_{Ap}\, x_{Aq}}{|\vec{x}_A|^5} -
\frac{\delta_{pq}}{|\vec{x}_A|^3} \right) 
+ {\cal O}(|\vec{x}|)\,.
\end{eqnarray}
Near $|\vec{x}|\approx 0$ one finds that, to first order in $b$,
$R(\sigma)_p$  is given by
\begin{equation}
  R(\sigma)_p = {3 \cdot 2^{19}\,b\over \chi\, q_1^{\,3}}
\;\frac{\delta_{pq} + \hat x_p \hat x_q}{|\vec{x}|} \;  
\sum_{A \neq 1} q_A \,   
\, \frac{(x_{A})^q }{|\vec{x}_A|^3} + {\cal O}(b^2)\,,
\end{equation}
where $\hat x_p $ denote the components of the unit vector. 
Thus, $R(\sigma)_p$ exhibits only a $\vert \vec x\vert^{-1}$
singularity so that the Riemannn curvature near the center $A=1$ is not
affected. Hence, in this simple example we recover  the
usual ${\rm AdS}_2\times S^2$ geometry typical for extremal black
holes. \\[3mm]

\noindent
{\large \bf Acknowledgments:} This work is supported in part by the
European Commission RTN programme HPRN-CT-2000-00131.
\smallskip
\noindent


\end{document}